\documentclass[runningheads,a4paper]{llncs}
\usepackage{llncsdoc}
\usepackage{amssymb}
\usepackage{graphicx}

\usepackage{url}
\urldef{\mailsa}\path|davies@stanford.edu|

\begin{document}

\mainmatter  

\title{Digital Rights and Freedoms:\\
A Framework for Surveying Users and Analyzing Policies}

\titlerunning{Digital Rights and Freedoms: A Framework}

%
%
\author{Todd Davies%
\thanks{To appear as: Todd Davies, ``Digital Rights and Freedoms: A Framework for Surveying Users and Analyzing Policies,'' in Luca Maria Aiello and Daniel McFarland (eds.), {\it Social Informatics: Proceedings of the 6th International Conference (SocInfo 2014)}, Barcelona, November 10-13, 2014, Springer LNCS. I would like to thank Laila Chima for assistance with the survey process, and Nathan Tindall for assistance with data analysis. I would also like to thank Jerome Feldman, Karl Fogel, Lauren Gelman, Rufo Guerreschi, Kaliya Hamlin, Dorothy Kidd, Geert Lovink, Mike Mintz, Brendan O'Connor, Doug Kensing, Steve Zeltzer, and the students who have taken Symsys 201 at Stanford for helpful discussions prior to the writing of this paper. Research costs were funded by the School of Humanities and Sciences at Stanford University.}%
}
\authorrunning{Digital Rights and Freedoms: A Framework}

\institute{Symbolic Systems Program and\\
Center for the Study of Language and Information\\
Stanford University\\
Stanford, California 94305-2150 USA\\
\mailsa\\
\url{http://www.stanford.edu/~davies}}

%
%

\toctitle{Digital Rights and Freedoms}
\tocauthor{A Framework for Surveying Users and Analyzing Policies}
\maketitle

\begin{abstract}
Interest has been revived in the creation of a ``bill of rights'' for Internet users. This paper analyzes users' rights into ten broad principles, as a basis for assessing what users regard as important and for comparing different multi-issue Internet policy proposals. Stability of the principles is demonstrated in an experimental survey, which also shows that freedoms of users to participate in the design and coding of platforms appear to be viewed as inessential relative to other rights. An analysis of users' rights frameworks that have emerged over the past twenty years similarly shows that such proposals tend to leave out freedoms related to software platforms, as opposed to user data or public networks. Evaluating policy frameworks in a comparative analysis based on prior principles may help people to see what is missing and what is important as the future of the Internet continues to be debated.
\end{abstract}

\section{Introduction}


In March of 2014, on the 25th anniversary of the proposal that led to the World Wide Web, its author Tim Berners-Lee launched an initiative called the Web We Want campaign, which calls for ``a global movement to defend, claim, and change the future of the Web'' \cite{WebWeWant}. The object of the campaign is an online ``Magna Carta,'' ``global constitution,'' or ``bill of rights'' for the Web and its users, which Berners-Lee argued was needed because ``the web had come under increasing attack from governments and corporate influence and that new rules were needed to protect the 'open, neutral' system'' \cite{Kiss}. 

Although the weight of Berners-Lee's voice in calling for a users' ``bill of rights'' is a recent development, the idea of a comprehensive user rights framework has been floated by others previously (see section 3 below). With more limited scope, over the past three decades, many initiatives have emerged to promote particular rights, abilities, and influence for users over their online environments and data. Both codified and informal concepts such as Free Software \cite{Stallman}, participatory design \cite{Kensing}, Open Source software \cite{Parens}, Creative Commons \cite{CC} and free culture \cite{Lessig}, data portability \cite{DPP}, and the DNT (Do Not Track) header \cite{DNT} are attempts to establish and promote principles outside of public policy through which people can participate in the decisions that affect them as software users. Other concepts, such as the right to connect \cite{RTC} and net neutrality \cite{NN} represent attempts to protect user rights and free access through public policy. This paper describes a broad set of principles guiding user freedom and participation, and relates these principles to past and ongoing initiatives introduced by others. 

\section{Rights, Freedoms, and Participation Principles}

We can analyze users' rights with reference to ten principles, which might be present (or not) to differing degrees in a particular software environment or policy framework. The principles below outline a framework for {\itshape opinion assessment} and {\itshape comparative analysis} rather than being intended as a policy proposal. It is important to keep in mind this distinction for what follows. The principles and the concepts defined in relation to them below were derived empirically from users' rights policy proposals, but they are not meant to be exhaustive in any sense.\footnote{Principles and concepts that lack citations in this section are referenced in the documents analyzed in Table 3, section 4.1.} 

\subsection{User Data Freedoms}

The first six of the principles (1-6) are amenable to adoption within a particular software platform or environment, which may be under either private or public ownership. Of these, the first three (1-3) pertain to the data generated by a given user, which are referred to in these descriptions as ``their data.''

\spnewtheorem{principle}{Principle}{\bfseries}{\rmfamily}

\begin{principle} {\itshape Privacy control.} The user is able to know and to control who else can access their data. \end{principle}

Some or all of the following concepts might appear in a privacy control policy.

\begin{alpherate}
\item
{\itshape Originator-discretionary reading control.} The user who generates data is able to read and to determine who else can read their data, and under what circumstances, and cannot have this ability taken away. Generated data may be created by the user deliberately, e.g. by filling out a form online or by posting a photograph, or it may be created as a byproduct of the user's behavior, such as click stream data or cookies from the user's browser that are read and stored on a site which they use. ({\itshape Do Not Track} initiatives are attempts to provide users with a partial form of this type of control \cite{DNT}.)

\item
{\itshape Data use transparency.} All policies and practices concerning the storage or transfer of a user's data are fully disclosed to the user prior to when the data are generated. This includes policies and practices of the software platform provider regarding the manner and length of time the user's data are stored. 

\item
{\itshape Usable privacy.} Access and control by a user of their data is practically feasible for the user. Access should be straightforward enough to be practical, and privacy settings should be as clear and easy to use as possible, including for novice users \cite{Karat2006}.

\item
{\itshape Nonretention of data.} User data are not retained without the consent of the user.
\end{alpherate}

\begin{principle} {\itshape Data Portability.} The user is able to obtain their data and to transfer it to, or substitute data stored on, a compatible platform. \end{principle}

Some or all of the following concepts might appear in a Data Portability policy, as defined by the Data Portability Project \cite{DPP}.

\begin{alpherate}
\item
{\itshape Free data access.} The user of a software platform is able to (a) download or copy all of their data, (b) download or copy all of the other data on the platform to which the user has access, and (c) know where their data are being stored, i.e. in what real world location or legal jurisdiction. 

\item
{\itshape Open formats.} The information necessary in order to read, interpret, and transfer data, i.e. application programming interfaces (APIs), data models, and data standards, are available to any user and are well documented.

\item
{\itshape Platform independence.} The user is able to access data while using a software platform independently of whether those data are stored within the platform or outside it in a compatible platform. Principles put forward in Data Portability policies that elaborate on this concept include the ability of the user to (a) authenticate or log in under an existing identity on another platform, (b) use data stored on another platform, (c) update their data on another platform and have the updates reflected in the platform in current use, (d) update their data on other platforms automatically by undating them on the platform in use, (e) share data stored on the platform in use with other platforms, and (f) specify the location or jurisdiction of storage for their data within the platform \cite{DPMP}.

\item 
{\itshape Free deletion.} The user can delete their account and all of their data, and these data will be removed or erased from storage in that platform consistent with the meaning of a transparently provided definition of deletion.
\end{alpherate}

\begin{principle} {\itshape Creative control.} The user is able to modify their data within the software platform being used, and to control who else can do so. \end{principle}

Some or all of the following concepts might appear in a creative control policy.

\begin{alpherate}
\item
{\itshape Originator-discretionary editing control.} Subject to transparency requirements, the user who generates data is able to edit and to determine who else can edit their data, and under what circumstances, and cannot have this ability taken away. Transparency requirements such as visibly maintaining past versions and making their existence apparent to any user who can access a data item are safeguards against the abuse of editing, which could otherwise be used to alter the historical record. 

\item
{\itshape Authorial copyright support.} The creator of content holds any legally allowed copyright over their data, and a user has the ability to prevent others who have access to their generated data from copying it for access by a third party who lacks access to the original. (This definition reflects an adaptation of traditional copyright for digital content, applying the Fair Use exemption to copying for private viewing by a party who already has authorized access.)

\item
{\it Reciprocal data sharing.} The user has the ability to permit people to copy (and possibly modify) their data for viewing by third parties subject to provisos such as the Attribution, Noncommercial, Share Alike, and No Derivatives requirements which can be imposed on the copying party in a {\itshape Creative Commons} license \cite{CC}.
\end{alpherate}

\subsection{Software Platform Freedoms}

Principles 4-6 pertain to the software platform in which users' data are created, edited, stored, and accessed. The descriptions of these principles distinguish different ways in which users may be able to participate in controlling, designing, and governing the operation of the software platform they use. 

\begin{principle} {\itshape Software freedom.} The user is able to modify code in the software platform being used, subject to rights of other users to control their own experience of the platform. \end{principle}

Some or all of the following concepts might appear in a software freedom policy.

\begin{alpherate}
\item
{\itshape Open Source code.} The source code that operates the software platform can be legally read, copied, downloaded, and modified by any user. Source code includes all the code that is necessary to operate the platform and to serve data to the user. (Open Source software is defined in the Open Source Definition \cite{Parens}.)

\item
{\itshape Reciprocal code openness.} The user and anyone else who modifies the platform's source code for use by others is legally bound to make their modified code available under licensing terms consistent with those under which the user legally accesses the source code. (This incorporates (a) the so-called ``copyleft'' provision of {\itshape Free Software} licenses such as the General Public License v.3, which require reciprocal sharing by anyone who distributes modified copies of the software to others in executable form \cite{Stallman}, and sometimes also (b) the Affero clause in the Affero General Public License v.3, which requires that code modifications be reciprocally shared by anyone who executes their modified version of the source code in a networked environment (e.g. over the Web) for use by others \cite{Affero}.)

\item
{\itshape User modifiable platform.} The user of a software platform has the ability to modify the code on the platform they are using, as long as doing so does not interfere with the rights of other users to experience the platform and interact with their data as they desire. (In its full form, this is a demanding provision that is not usually satisfied in practical platforms, though it is often fulfilled in limited ways, e.g. by permitting a user-selected, configurable interface. This concept is an extension of the ideas in \cite{Stallman} to networked platforms.)
\end{alpherate}

Principles 1-4 are freedoms of individual users, which can be composed to define freedom for a community of users. Principles 5-6 are defined at the level of the group of users of a given software platform, which for each individual user means the freedom to participate in a collective process that determines the design and governance of a software and data environment. These last two freedoms allow for an especially large range of freedoms and participation mechanisms.

\begin{principle} {\itshape Participatory design.} The design of the platform is produced by all of its users. \end{principle}

Some or all of the following concepts might appear in a participatory design policy \cite{Kensing,SchulerNamioka}.

\begin{alpherate}
\item
{\itshape User-centered design.} The needs and desires of users are the primary or sole factor driving the design. Users' needs may be assessed in various ways, e.g. through ethnographic observation, surveys, one-on-one interviews, and focus groups, that focus on the problems and goals of users at a functional level.

\item
{\itshape User input to design.} The users' preferences and beliefs about design choices are collected and influence the design of the platform. This type of input can include, for example, expressions of preference between different options that are presented by a designer.

\item
{\itshape User-generated design.} Users participate in the creation of design solutions as actual partners in the design team, e.g. providing ideas through brainstorming with designers and/or other users, and helping to solve design problems creatively.

\item
{\itshape Customizable design.} Users can individually or collectively redo or configure parts of the platform's design and this feature is itself part of the design.
\end{alpherate}

\begin{principle} {\itshape User self-governance.} The operation of the platform is governed by all of its users. \end{principle}

Some or all of the following concepts might appear in a user self-governance policy. Wikipedia self-governance implements all of these concepts in varying degrees \cite{WikipediaGovernance}.

\begin{alpherate}
\item
{\itshape Participatory policy making.} Users are involved in creating and making decisions about the framework of rules and practices governing the platform they are using. (This can range from input on proposed policies, to voting, to full-fledged deliberative democracy online \cite{Davies-Gangadharan}.) 

\item
{\itshape Participatory implementation.} Users are involved in executing and enforcing the policies that govern the platform. (This can include forms of participation such as monitoring one's own compliance with policies, notifying other users of policy violations, raising and discussing implementation questions in online forums, and serving in defined roles.) 

\item
{\itshape Participatory adjudication.} Users are involved in making judgments when human judgment (usually in a collective form) is a part of the platform's operation, e.g when a policy implementation question is in dispute. when content much be judged appropriate or not under defined criteria or procedures, or when the platform asks users for input in rendering a judgment or rating concerning user content. 
\end{alpherate}

\subsection{Public Network Freedoms}

Principles 7-10 generally require public policy adoption, such as legislation, executive orders, or international agreements.

\begin{principle} {\itshape Universal network access.} Every person is legally and practically able, to the greatest extent possible, to access the Internet, and it is available everywhere in a form adequate for both retrieving and posting data. \end{principle}

Some or all of the following concepts might appear in a universal network access policy.

\begin{alpherate}
\item
{\itshape Right to connect.} Internet access cannot be denied to a user or to a population of users wherever it is possible to provide access \cite{RTC}.

\item
{\itshape Universal digital literacy.} Every person who possesses the intellectual ability to do so develops the skills to use the Internet as both a recipient and producer of information, to the maximal achievable for meaningful individual participation in a democracy.

\item
{\itshape No- or low-cost service.} Cost is not a barrier to accessing the Internet.

\item
{\itshape Omnipresent service.} The Internet is available everywhere and at all times.

\item
{\itshape Accessibility.} Internet access is available to everyone in a way that matches their physical and mental abilities.
\end{alpherate}

\begin{principle} {\itshape Freedom of information.} Every person is legally and practically able to produce and receive information in the way that they want, to the maximal extent consistent with the rights of others. \end{principle}

Some or all of the following concepts might appear in a freedom of information policy.

\begin{alpherate}
\item
{\itshape Right to privacy.} Private communications cannot be intercepted, monitored, or stored by governments or other entities without due process to establish a compelling public interest.

\item
{\itshape Right to anonymous speech.} Everyone is able to both receive and produce public information without being required to identify themselves, either implicitly or explicitly.

\item
{\itshape Freedom from censorship.} Free expression, without political restrictions, is protected both for producers and receivers of information.

\item
{\itshape Open Access to all publicly funded data.} Government data and that which is produced through publicly funded research is available freely to everyone. 

\item
{\itshape Democratically controlled security.} Government security policies must be as transparent as possible to allow for them to be publicly debated, and those who oversee them must be accountable to everyone.

\item
{\itshape Right to be forgotten.} Everyone is able to have information about them made inaccessible to others when these data are determined by established procedures to be either no longer relevant or unfairly stigmatizing to their subject(s) \cite{Rosen}.
\end{alpherate}

\begin{principle} {\itshape Net neutrality.} All providers of Internet connections and services are legally and practically required to treat data equally as it is transmitted through the infrastructure they control. \end{principle}

Some or all of the following concepts might appear in a net neutrality policy. Disallowed forms of discrimination against data would include blocking data or charging fees in exchange for allowing it to be transmitted \cite{NN}.

\begin{alpherate}
\item
{\itshape Source neutrality.} Providers of network connections may not discriminate against data on the basis of its origin, e.g. another service provider or a particular social media platform.

\item
{\itshape Format neutrality.} Providers of network connections may not discriminate against data on the basis of its format, e.g. MIME type, protocol, or port.

\item
{\itshape Content neutrality.} Providers of network connections may not discriminate against data on the basis of its content, e.g. political expression with which the provider disagrees.

\item
{\itshape End-user neutrality.} Providers of network connections may not discriminate against data on the basis of the end user's identity. 
\end{alpherate}

\begin{principle} {\itshape Pluralistic open infrastructure.} Everyone has access to multiple independent but interoperating software platforms as options for their data. \end{principle}

Some or all of the following concepts might appear in a pluralistic open infrastructure policy.

\begin{alpherate}
\item
{\itshape Multiplicity of platforms.} Policies ensure that all users have multiple software platforms to choose from as environments for their data.

\item
{\itshape Decentralized control.} Software platforms are coordinated to interoperate in a way that is not controlled by any one government, authority, or interest.

\item
{\itshape Transparent control.} Common infrastructure and standards are developed and documented in a way that is open and understandable to anyone.
\end{alpherate}

\section{A Survey of Internet Users}

To illustrate the use of this analysis framework for surveying users, a demonstration survey was conducted using Amazon Mechanical Turk in the summer of 2014. 

\subsection{Participants and Method}

A total of 780 survey takers completed a survey on the Qualtrics platform\cite{Qualtrics}. Each survey taker was shown a subset of the principles and concepts described in the framework of section 2, and asked to ``rate [on a 0 to 10 scale, moved left or rigth from the midpoint] how important you think it is for the user of a software `platform' (such as a website, app, operating system, or social network)'' to have the particular right or freedom described in each statement they read.\footnote{Survey materials and data for this study are published on the Harvard Dataverse Network at \url{doi:10.7910/DVN/27510}.} The statements consisted of the unparenthesized and unbracketed portions of each principle and concept in section 2, with the title of each excluded. Participants were told: ``These statements describe what {\itshape could} be true in some situations or hypothetically, not necessarily what {\itshape is} true now or in some particular situation.'' 

The survey as designed assumed users were fluent in reading English, and able to understand digital concepts such as ``data'' and ``software platform.'' Participants were recruited on the Amazon Mechanical Turk platform\cite{AMT}, with a link to the survey on Qualtrics. The survey was open only to U.S.-located respondents whose prior approval percentage by requestors on MTurk exceeded 98\%. Participants were 39\% female and 61\% male. Respondents' reported age groups were 3\% under 20, 41\% 20-29, 33\% 30-39, 12\% 40-49, 8\% 50-59, 3\% 60-69, 0.4\% 70-79, and 0\% over 80. Thirteen percent reported being ``very knowledgeable about digital rights and freedoms,'' while 72\% reported being ``somewhat knowledgeable'' and 15\% ``not knowledgeable.'' 

The participant pool, while not representative of the population of the United States as a whole (let alone the world), nonetheless represents a population of interest: relatively sophisticated users who could be expected to have heard of at least some of the concepts in our framework. The intent was both to test whether users would have consistent views of these statements, and to assess relative levels of support for the principles and concepts in the framework within the young-skewing demographic of high functioning Internet users. Although we will not do it here, the gender, age group, and knowledge data could be used to adjust for sampling bias relative to the general population of users in the United States. (A more complete demographic analysis is planned in a future paper.)

For this survey, participants randomly saw either a {\itshape broad} rating set consisting of a random ordering of all ten of the primary principles in the framework (a {\itshape within}-group comparison of the ten principles), or a {\itshape narrow} random ordering of a subset of principles and concepts that included one of the primary principles and its associated concepts (a {\itshape between}-groups comparison of the ten principles). This allowed for cross-item comparative and correlational analysis for both the ten principles and for the concepts associated with each principle along with that principle itself. Random assignment of participants into the broad or narrow rating sets created an experiment for testing whether average importance ratings for the ten principles would remain stable across these two rating contexts.

\subsection{Survey Results}

\begin{table}
\centering
\caption{Importance Ratings of Principles and Concepts (Narrow Rating Sets)}
\begin{tabular}{lccc}
\hline
Principle/Concept & Mean (0-10) & N & Std. Dev. \\
\hline
{\bfseries 1--Privacy control} & 8.69 & 71 & 1.9  \\
\ 1a Originator-discretionary reading control & 7.96 & 71 & 2.2  \\
\ 1b Data use transparency & 8.06 & 71 & 2.2   \\
\ 1c Usable privacy & 8.58 & 71 & 1.7   \\
\ 1d Nonretention of data & 8.65 & 71 & 1.7 \\
{\bfseries 2--Data Portability} & 7.90 & 69 & 1.7 \\
\ 2a Free data access & 7.74 & 69 & 2.0 \\
\ 2b Open formats & 6.61 & 68 & 2.5 \\
\ 2c Platform independence & 6.71 & 68 & 2.3 \\
\ 2d Free deletion & 8.96 & 69 & 1.5 \\
{\bfseries 3--Creative control} & 7.77 & 65 & 2.4 \\
\ 3a Originator-discretionary editing control & 7.36 & 66 & 2.3 \\
\ 3b Authorial copyright support & 7.65 & 66 & 2.5  \\
\ 3c Reciprocal data sharing & 6.64 & 65 & 2.5 \\
{\bfseries 4--Software freedom} & 6.01 & 73 & 2.7 \\
\ 4a Open Source code & 5.85 & 71 & 2.7 \\
\ 4b Reciprocal code openness & 5.52 & 72 & 2.6 \\
\ 4c User modifiable platform & 6.63 & 74 & 2.6 \\
{\bfseries 5--Participatory design} & 5.48 & 77 & 2.4 \\
\ 5a User-centered design & 7.08 & 75 & 2.1 \\
\ 5b User input to design & 6.83 & 77 & 1.9 \\
\ 5c User-generated design & 6.16 & 76 & 2.5 \\
\ 5d Customizable design & 6.36 & 77 & 2.2 \\
{\bfseries 6--User self-governance} & 5.82 & 63 & 2.7 \\
\ 6a Participatory policy making & 6.30 & 64 & 2.5 \\
\ 6b Participatory implementation & 6.01 & 66 & 2.6 \\
\ 6c Participatory adjudication & \dag &  &  \\
{\bfseries 7--Universal network access} & 8.40 & 82 & 1.9 \\
\ 7a Right to connect & 8.56 & 82 & 1.7 \\
\ 7b Universal digital literacy & 7.45 & 82 & 2.2 \\
\ 7c No- or low-cost service & 8.27 & 82 & 2.2 \\
\ 7d Omnipresent service & 8.43 & 82 & 2.0 \\
\ 7e Accessibility & 7.12 & 80 & 2.8 \\
{\bfseries 8--Freedom of information} & 8.01 & 74 & 1.7 \\
\ 8a Right to privacy & 8.72 & 74 & 1.9 \\
\ 8b Right to anonymous speech & 7.39 & 74 & 2.3 \\
\ 8c Freedom from censorship & 8.46 & 74 & 1.8 \\
\ 8d Open Access to publicly funded data & 8.20 & 74 & 2.0 \\
\ 8e Democratically controlled security & 8.12 & 74 & 1.9 \\
\ 8f Right to be forgotten & 7.59 & 74 & 2.3 \\
{\bfseries 9--Net neutrality} & 8.11 & 61 & 2.5 \\
\ 9a Source neutrality & 8.39 & 61 & 2.2 \\
\ 9b Format neutrality & 7.42 & 61 & 2.5 \\
\ 9c Content neutrality & 8.56 & 61 & 2.2 \\
\ 9d End-user neutrality & 8.52 & 61 & 2.2 \\
{\bfseries 10--Pluralistic open infrastructure} & 6.84 & 70 & 2.1 \\
\ 10a Multiplicity of platforms & 7.24 & 70 & 2.0 \\
\ 10b Decentralized control & 8.27 & 70 & 1.8 \\
\ 10c Transparent control & 7.97 & 70 & 1.8 \\
\hline
\multicolumn{4}{l}{\footnotesize \dag \ Ratings for 6c were not meaningful: incorrect wording on survey.}
\end{tabular}
\end{table}

The mean importance ratings from the narrow sets for each principle and concept, together with sample sizes and standard deviations, are shown in Table 1. Standard errors ranged from 0.2 to 0.4 and are easily calculated from the table.  As can be seen from Table 1, highly rated primary principles tend to have highly rated associated concepts, but there are occasional deviations within principle-concept groupings. The concepts associated with data portability, for example, ranged widely in support, from a 6.61 rating for open formats (2b) to an 8.96 rating for free deletion (2d). Pluralistic open infrastructure, as worded in the principle, drew less support (6.84) than any of its three associated concept statements, which ranged from 7.24 to 8.27. Every principle and concept was rated significantly above the midpoint (and starting point) of 5.0 in this survey, indicating that participants on average regarded each of them as at least somewhat important. 

\begin{table}
\centering
\caption{Comparing Importance Ratings of the Ten Principles}
\begin{tabular}{ccccccccccccc}
\hline
Principle \ & Broad & Narrow & Aggregate \ & \multicolumn{9}{c}{Correlations of Importance Ratings (Broad Set)} \\
Number & Mean & Mean & Mean & {\bfseries 2} & {\bfseries 3} & {\bfseries 4} & {\bfseries 5} & {\bfseries 6} & {\bfseries 7} & {\bfseries 8} & {\bfseries 9} & {\bfseries 10} \\
\hline
{\bfseries 1} & 9.09 & 8.69 & 8.89 & 0.50\ddag & 0.48\ddag & 0.02 & -0.06 & 0.24 & 0.40\ddag & 0.66\ddag & 0.15 & 0.06 \\
{\bfseries 2} & 7.74 & 7.90 & 7.82 &     & 0.35\ddag & 0.08 & 0.05 & 0.31\dag & 0.51\ddag & 0.66\ddag & 0.35\ddag & 0.36\ddag \\
{\bfseries 3} & 7.86 & 7.77 & 7.82 &     &     & 0.19 & 0.07 & 0.33\dag & 0.22 & 0.48\ddag & 0.37\ddag & 0.13 \\
{\bfseries 4} & 5.55 & 6.01 & 5.78 &     &     &     & 0.36\ddag & 0.22 & 0.14 & 0.18 & 0.05 & 0.31\dag \\
{\bfseries 5} & 5.12 & 5.48 & 5.30 &     &     &     &     & 0.55\ddag & -0.02 & 0.18 & 0.28\dag & 0.50\ddag \\
{\bfseries 6} & 6.05 & 5.82 & 5.93 &     &     &     &     &     & 0.25\dag & 0.37\ddag & 0.49\ddag & 0.43\ddag \\
{\bfseries 7} & 8.58 & 8.40 & 8.49 &     &     &     &     &     &     & 0.48\ddag & 0.26\dag & 0.27\dag \\
{\bfseries 8} & 7.86 & 8.01 & 7.94 &     &     &     &     &     &     &     & 0.36\ddag & 0.42\ddag \\
{\bfseries 9} & 8.02 & 8.11 & 8.06 &     &     &     &     &     &     &     &     & 0.43\ddag \\
{\bfseries 10} & 6.55 & 6.84 & 6.69 &     &     &     &     &     &     &     &     &     \\
\hline
& & & & \multicolumn{9}{l}{\footnotesize \dag \ denotes $p < .05$, and \ddag \ denotes $p < .005$.}
\end{tabular}
\end{table}

A full analysis of all of the principle-concept groups is beyond what we have room for in this paper, but Table 2 shows the basis for such an analysis of the ten primary principles. This table displays mean ratings first for the broad rating set -- participants who rated all and only the primary principles -- and compares them to the narrow set means. The aggregage means are simply the averages of the broad and narrow means. The correlation between the means of the within- and between-groups surveys is extremely high (.98), indicating that attitudes toward the principles are stable across these two different presentation contexts for this population. The most important primary principle in the eyes of participants was the statement that is labeled ``privacy control'' in section 2 (agg. mean 8.89), though again participants did not see the labels. Next highest were 7-universal access (8.49), 9-net neutrality (8.06), 8-freedom of information (7.94), and 2-data portability and 3-creative control (both 7.82). The remaining principles formed a less highly rated cluster: 10-pluralistic open infrastructure (6.69), 6-user self-governance (5.93), 4-software freedom (5.78), and 5-participatory design (5.30).  

Table 2 also shows correlations between the importance ratings of pairs of principles for participants in the broad rating set condition: those who rated all ten of the main principles instead of just one. All of the significant correlations, and most of the nonsignificant ones, are positive, indicating a general disposition for individuals to be more or less favorable to digital rights and freedoms. Ratings for the lowest rated principles (4 and 5) were significantly correlated, but ratings between principle 4 or 5 and the other principles tended not to be significant. In the set of correlations involving just one of principles 4 and 5, only 3 out of 14 were significant, whereas in the remaining correlations, 25 out of 30 were significant. Consistent with their low overall average ratings, this indicates that principles 4 and 5 are evaluated differently by users compared to the rest of the principles ($p=.0001$ by a Fisher exact test). 

\subsection{Survey Lessons}

The use of the framework in this survey has demonstrated that it is possible to obtain meaningful results about the relative importance that users attach to different digital rights and freedoms. Meaningfulness in this case is demonstrated by the nearly perfect consistency between average ratings in two different contexts. In the narrow rating set (between-groups rating of principles), participants considered only one primary principle and several other concepts that were chosen for their close relationship to the primary principle. In the broad set (within-group rating of principles), participants saw all of the principles. These different contexts might have been thought to influence respondents differently. In terms of average ratings, that does not appear to happen in this population. 

A second finding, which we can see in Table 2, is that while most of the principles tend to be significantly correlated with each other, indicating that people who tend to favor users' rights under one principle tend to favor them under other principles, there are exceptions to this pattern. The tendency of a user to favor privacy control, data portability, or universal network access (all of which are highly correlated with each other) is not predictive of a high rating for participatory design or software freedom. Indeed, in the narrow rating set, the principles fell into two groupings, and the lowest rated principles were those most associated with user participation in the software environment.  

\section{Users' Rights Frameworks} 

To illustrate the application of the principles to policy, we will analyze four policy frameworks that have been proposed over the past twenty years aimed at securing rights for users:\footnote{For other users' rights framework proposals, see \cite{Curtis,Karat,OReilly,Patrianakos}.} (1) {\itshape Rights and Responsibilities of Electronic Learners} (RREL, 1994). An early framework was developed as part of this project within the American Association for Higher Education (AAHE), after extensive input from the education community, and was described by American University computer science professor Frank W. Connolly \cite{Connolly}.\footnote{An earlier paper laying out the motivations and a procedure for drafting such a document was published in 1990 by Connolly, Gilbert, and Lyman \cite{ConnollyEtAl}.} (2) {\itshape A Bill of Rights for Users of the Social Web} (BRUSW, 2007). Social media engineer Joseph Smarr and colleagues \cite{Smarr} delineated a set of ``fundamental rights'' to which ``all users of the social web are entitled.'' (3) {\itshape Marco Civil da Internet} (MCdI, 2014). In recent years, Brazil has taken the lead in initiatives to define a ``constitution of the Internet.'' In March and April, 2014, Brazil's two legislative chambers each passed the {\itshape Marco Civil da Internet} (Civil Rights Framework for the Internet). The priority placed on the {\itshape Marco Civil} followed a 2013 United Nations speech by the country's president, Dilma Rousseff, who ``presented proposals for a civilian multilateral framework for the governance and use of the Internet, capable of ensuring such principles as freedom of expression, privacy of the individual and respect for human rights, as well as the construction of inclusive and non-discriminatory societies'' \cite{Rousseff,Marco}.\footnote{Reportedly, Article 12 was struck from the draft version before final passage \cite{Mari}.} (4) {\itshape NETmundial Draft Outcome Document} (NDOD, 2014). In its international role as a leader in recent Internet governance initiatives, Brazil was the host of the Global Multistakeholder Meeting on the Future of Internet Governance, also known as NETmundial, in April 2014. President Rousseff announced the meeting in October 2013, after revelations that the U.S. National Security Agency had monitored her phone calls and email messages \cite{Kelion}, and the Draft Outcome Document was posted on the Web for open comment on April 14 \cite{IGP}.\footnote{The Draft Document was refined on April 24, 2014, into a ''Multistakholder Statement,'' \cite{NMMS} but the draft document is used here because it is annotated with section references for easier analysis.}

\subsection{Comparison of Frameworks}

\begin{table}
\caption{Analysis of Four Users' Rights Frameworks (see footnotes on previous page)}
\begin{tabular}{llllll}
\hline\noalign{\smallskip}
Principle/Concept & RREL\footnotemark & BRUSW\footnotemark & MCdI\footnotemark & NDOD\footnotemark \\
\noalign{\smallskip}
\hline
\noalign{\smallskip}
{\bfseries 1--Privacy control} & & & & \\
\ 1a Originator-discretionary reading control & I:3,IV:4 & a,b & 7:VII,8,10 & \\
\ 1b Data use transparency & I:3 & b$\ast$ & 7 & \\
\ 1c Usable privacy & & & & \\
\ 1d Nonretention of data & & & 15$\ast$,16$\ast$,17$\ast$ & \\
{\bfseries 2--Data Portability} & & & & \\
\ 2a Free data access & I:5$\ast$ & a$\ast$,c$\ast$ & & \\
\ 2b Open formats & & i & & \\
\ 2c Platform independence & & c,i,ii,iii & & \\
\ 2d Free deletion & I:5$\ast$ & b$\ast$ & 7:X & \\
{\bfseries 3--Creative control} & & & & \\
\ 3a Originator-discretionary editing control & I:3,I:5 & & & \\
\ 3b Authorial copyright support & I:5 & & 20$\ast$ & \\
\ 3c Reciprocal data sharing & & & & \\
{\bfseries 4--Software freedom} & & & & \\
\ 4a Open Source code & & & & \\
\ 4b Reciprocal code openness & & & & \\
\ 4c User modifiable platform & & & & \\
{\bfseries 5--Participatory design} & & & & \\
\ 5a User-centered design & & & & \\
\ 5b User input to design & & & & \\
\ 5c User-generated design & & & & \\
\ 5d Customizable design & & & & \\
{\bfseries 6--User self-governance} & & & & \\
\ 6a Participatory policy making & & & & \\
\ 6b Participatory implementation & & & & \\
\ 6c Participatory adjudication & & & & \\
{\bfseries 7--Universal network access} & & & & \\
\ 7a Right to connect & I:1,IV:1 & & 7:III & 7,23 \\
\ 7b Universal digital literacy & I:2 & & 7:XI,19:VIII,27 & 23 \\
\ 7c No- or low-cost service & & & & 23 \\
\ 7d Omnipresent service & & & 7:IV & 10,11 \\
\ 7e Accessibility & & & 25 & 6,23 \\
{\bfseries 8--Freedom of information} & & & & \\
\ 8a Right to privacy & I:3,IV:2 & & 7,8,10,11 & 5 \\
\ 8b Right to anonymous speech & I:4 & & & \\
\ 8c Freedom from censorship & I:4 & & & 3 \\
\ 8d Open Access to publicly funded data & & & & \\
\ 8e Democratically controlled security & & & 10:IV$\ast$ & \\
\ 8f Right to be forgotten & & & & \\
{\bfseries 9--Net neutrality} & & & & \\
\ 9a Source neutrality & & & 9 & 12$\ast$ \\
\ 9b Format neutrality & & & 9 & 12$\ast$ \\
\ 9c Content neutrality & & & 9 & 12$\ast$ \\
\ 9d End-user neutrality & & & 9 & 12$\ast$ \\
{\bfseries 10--Pluralistic open infrastructure} & & & & \\
\ 10a Multiplicity of platforms & III:1 & & 19:VII+X & 11 \\
\ 10b Decentralized control & III:3 & & 19:I-VI & 13,15,16,19-22,24 \\
\ 10c Transparent control & & & & 17,25 \\
\hline
\end{tabular}
\end{table}
\addtocounter{footnote}{-3}\footnotetext{Locations are coded as Article:Section as seen in \cite{Connolly}.}
\addtocounter{footnote}{1}\footnotetext{Locations are coded according to the inserted letters and roman numerals in the description of this framework above.}
\addtocounter{footnote}{1}\footnotetext{The {\itshape Marco Civil} applies only to Brazil, so the public freedoms (Principles 7-10) must be understood in that light. Locations are coded as Article[:Section] as seen in \cite{Marco}.}
\addtocounter{footnote}{1}\footnotetext{Locations are coded by paragraph number, as shown at \url{http://document.netmundial.br/1-internet-governance-principles/}.}

An analysis based on the framework of section 2 of each of the four texts yields the results in Table 3, where a location reference means that the principle or concept is clearly and substantially present (explicitly or implied) in the text, in a positive way (meaning that the concept is affirmed as a right); a blank entry means it is apparently not present; and a location reference followed by an asterisk ``$\ast$'' indicates ambiguity about whether the concept is present or not. 


The table shows firstly that none of the frameworks covers all of the principles. But some are more comprehensive than others. While the RREL and MCdI frameworks span concepts in both the user data freedoms (principles 1-3) and public network freedoms (principles 7-10), the BRUSW and NDOD frameworks are more specialized. The BRUSW framework was put forward as a set of rights for social Web users, and is limited to user data freedoms. The NDOD framework, on the other hand, is a global Internet governance initiative that seeks only to regulate at the international level. RREL and MCdI span two regions of the table for different reasons. RREL was an early and somewhat more vague attempt to establish principles that might apply either to public policy or to users of a specific platform. MCdI, on the other hand, is a draft law for a specific jurisdiction (Brazil) with authority to regulate software platforms that are subject to the country's laws, so that it may limit the freedom of platforms in the course of regulating at a national level. 

None of the four frameworks analyzed above (or the additional ones referenced in footnote 3) include provisions that appear to enact what are herein called software platform freedoms (principles 4-6). It appears, from these data, that giving users power over their software environment, through software freedom, participatory design, and user self-governance, are not strong values among those who have constructed these frameworks. These were also the lowest rated three principles in the survey reported in section 3.

\subsection{Benefits of an Analysis Framework}

The principles and concepts of section 2 comprise an analysis framework, as opposed to the policy frameworks analyzed in Table 3. An analysis framework of this kind gives us the following types of leverage for understanding users' rights policies: (a) it allows for easier comparison across frameworks; (b) it allows us to see what is missing from a {\itshape particular} policy framework; (c) it facilitates further study of the dimensions that characterize users' rights, e.g. surveys of users and policy makers to determine strengths of priority for different freedoms; and (d) it allows us to see persistent gaps {\itshape across} policy frameworks, such as the apparent lack of attention to software platform freedoms (principles 4-6).

\section{Conclusion}

The current moment is one of revival for the idea of a ``bill of rights'' or ``constitution'' for users online. On one hand, some observers have expressed skepticism about the feasibility of this concept, particularly at the International level (e.g. \cite{Clark}). On the other hand, the growing control as well as documented instances of misuse of power by governments appear to have fed this new level of interest on the part of figures such as Tim Berners-Lee and Dilma Rousseff. Whether this will translate into lasting change remains to be seen. But there remain many levels at which policies can be adopted, from particular software platforms and small online communities to the entire world.

It seems likely that discussions about the principles that govern Internet users will continue to pick up steam in the years ahead, and, if the history of other major shifts in civilization is a guide, the process will lag technological change considerably. If many people think carefully about the principles they want to govern their own use of the Internet, articulate those principles, and invoke them in discussing policy proposals, we may have a better chance of arriving at arrangements that satisfy most users and that meet the needs of contemporary societies.

\end{document}